\documentclass[10pt,a4paper,twocolumn]{article}

\usepackage[labelfont=bf,font=small,margin=10pt]{caption}
\usepackage{upgreek}
\usepackage[margin=20mm]{geometry}
\usepackage{pgfplots}
\pgfplotsset{compat=1.14}

\newcommand{\ignore}[1]{}

\newcommand{\rp}{\ensuremath{r_\mathrm{p}}}
\newcommand{\rt}{\ensuremath{r_\mathrm{t}}}
\newcommand{\rti}{\ensuremath{r_{i}}}
\newcommand{\Vp}{\ensuremath{V_\mathrm{p}}}

\newcommand{\Vtot}{\ensuremath{V_\mathrm{0}}}

\newcommand{\mrm}[2]{\ensuremath{#1_\mathrm{#2}}}
\newcommand{\mr}[1]{\ensuremath{\mathrm{#1}}}

\newcommand{\average}[1]{\ensuremath{\langle#1\rangle}}
\newcommand{\req}[1]{(\ref{eq:#1})}
\newcommand{\rep}[1]{\ref{fig:#1}}
\newcommand{\ret}[1]{\ref{tab:#1}}
\newcommand{\dg}[1]{\ensuremath{#1^\circ}}
\newcommand{\latitude}{\ensuremath{\Lambda}}

\newcommand{\X}{\ensuremath{{\mathrm{X}}}}
\newcommand{\Xp}{\ensuremath{{\mathrm{p}^{+}}}}
\newcommand{\Xpi}{\ensuremath{\mathrm{\uppi}}}
\newcommand{\Xpipm}{\ensuremath{\mathrm{\uppi}^{\pm}}}
\newcommand{\XK}{\ensuremath{\mathrm{K}}}
\newcommand{\XKpm}{\ensuremath{\mathrm{K}^{\pm}}}

\newcommand{\ptp}{\ensuremath{p_{\mr{pu}}}}

\title{
Thunder-cell as Source of Energetic Protons}
\author{Ale\v{s} Berkopec\footnote{contact: ales.berkopec@gmx.com}}
\date{September 29, 2020}

\begin{document}

\maketitle

\begin{abstract}
In this article we present the following hypothesis: thunder-cell
	ejects highly energetic protons, each of which
	creates a tree-structure of weakly ionized
	trajectories that can develop into a lightning
	channel. The tree-structure and the channel have the
	same geometry so the mean free path of a proton
	corresponds to the average length of the channel
	between two successive nodes (branching points). We
	show this length is around $660$~m in lower Earth
	atmosphere. Effects of Coulomb interaction and
	various outcomes of proton-nucleus reaction are
	taken into account. A prediction of CG/CC ratio that
	follows agrees well with the available data, but
	only measurements of lightning geometry can reveal
	whether the hypothesis is any closer to the correct
	explanation of the phenomenon.

\bigskip\noindent
{\slshape Keywords}: 
lightning initial phase, CG and CC lightning
\end{abstract}

\section{Introduction}

The reported values of potential gradient before and during
a lightning strike are at least one order of magnitude
smaller compared to the ones required to induce sparks under
controlled conditions in laboratories
\cite{Gunn1948,Marshall1991,Winn1974}. We tried to explain
this difference with theoretical prediction \cite{Berkopec2012} 
involving fast charged particles that precede stepped
leader. This idea was later encouraged by a report about
weak correlation between lightning frequency and solar wind
intensity \cite{Scott2014}. Since every CG (cloud-to-ground)
and CC (intra-cloud, inter-cloud or cloud-to-air) lightning
channel starts in a thunder-cell the intense freezing of
super-cooled water inside thunder-cell was proposed as a
process that might lead to ejection of charged projectiles
\cite{Berkopec2016}. 

In view of the hypothesis, trajectory of the primary
projectile and its collision products, all electrically
charged, determine the geometry of the stepped leader and
subsequent lightning channel. Interaction of the projectile
with electrons in air slows the projectile down and ionizes
the trajectory, while collisions with the nuclei may produce
higher-order projectiles and result in branching of the
channel. At the end of the process, weakly ionized tree
structure forms a conducting path between cloud and ground
(CG lightning, Fig.~\rep{lch}), or cloud and a point in the air
(CC lightning).

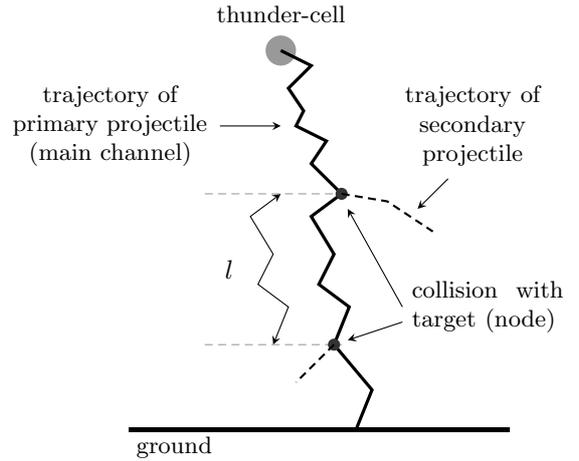
\begin{figure}[hbt]
\centerline{
\begin{tikzpicture}[scale=1.0]
\fill[black!100] (-2,0) -- (3,0) -- (3,-0.06) -- (-2,-0.06) -- cycle; 
\node[anchor=north] at (-1.4,0) {\small ground};
\fill[black!40] (0,5) circle (0.20);
\node[anchor=south] at (0.0,5.25) {\small thunder-cell};
\draw[stealth-stealth,thin] (0.0,3.1) -- (-0.4,2.8) -- (-0.1,2.3) -- (-0.3,1.9) -- (0.1,1.6) --
(-0.1,1.1);
\draw[densely dashed,black!40] (-1.0,3.1) -- (0.8,3.1);
\draw[densely dashed,black!40] (-1.0,1.1) -- (0.8,1.1);
\node[anchor=east] at (-0.5,2.1) {$l$};
\draw[very thick] (0,5) -- (0.4,4.8) -- (0.1,4.5) -- (0.3,4.2) --(0.2,4.0) --(0.6,3.8) --(0.4,3.5) -- (0.8,3.1) -- (0.4,2.8) -- (0.7,2.3) -- (0.5,1.9) -- (0.9,1.6) -- (0.7,1.1) -- (1.2,0.5) -- (1,0);
\fill[black!80] (0.8,3.1) circle (0.08);
\draw[densely dashed,thick] (0.8,3.1) -- (1.4,3.0) -- (2.0,2.6);
\fill[black!80] (0.7,1.1) circle (0.08);
\draw[densely dashed,thick] (0.7,1.1) -- (0.2,0.6);
\node[anchor=east] at (-0.72,4.0)
{\parbox{28mm}{\centering\small trajectory of primary
projectile (main channel)}};
\draw[-stealth] (-0.8,4.0) -- (0.0,4.0);
\node[anchor=west] at (1.2,4.0) {\parbox{24mm}{\centering\small trajectory of secondary projectile}};
\draw[-stealth] (2.2,3.3) -- (1.8,2.9);
\node[anchor=west] at (1.6,1.6) {\parbox{20mm}{\small
collision with target (node)}}; 
\draw[-stealth] (1.6,1.6) -- (0.88,2.9);
\draw[-stealth] (1.6,1.4) -- (0.88,1.16);
\end{tikzpicture}
}
\caption{Thunder-cell ejects a projectile. 
Why protons are best candidates for projectiles is explained
	in Section 2, for the estimation of length
	between two successive collisions $l$ see
	Sections 3 and 5, and for a role of Coulomb
	interaction see Section 4.} 
\label{fig:lch}
\end{figure}

The projectile and the ionized tree-structure it leaves
behind provide explanation for at least three aspects of CG
and CC lightnings: a) lower values of potential gradient
sufficient to initiate lightning, b) tree-structure of a
lightning channel with loosely defined direction, 
and c) dependence of CG/CC ratio on height of the cloud-base
and consequently on latitude.

\section{Projectile: type of particle}


We reason the projectile is expected to have the following
characteristics:


\begin{enumerate}
\setlength{\itemsep}{0pt}
\item[a)] {\slshape electrically charged}

Only a charged particle leaves behind an ionized trajectory.
		Coulomb interaction has only a minor effect on
		the geometry of the channel, as we show
		later.
\item[b)] {\slshape induces secondary projectiles of the
	same type}

Lightning channel is often split, however, the branches that
grow from the split point are indistinguishable.
\item[c)] {\slshape elementary particle}

Atomic nuclei of elements heavier than hydrogen or other
		composite particles do not fit the role.
		Their collisions with nuclei in the air
		would lead to a spallation at such energies
		\cite{Berkopec2012}, so one could observe
		lightning channels with also three or more
		branches continuing from a single node. Such
		channels have not been observed in CG and CC
		lightnings.
\end{enumerate}

The most suitable candidates for the role are protons: they
are electrically charged, they do not decay, pick-up
reactions $(\Xp,2\Xp)$ (proton-projectile hits a nucleus,
passes through, and ejects additional, secondary
proton-projectile) are common, and they are not composite.

Other types of particles, like pions and kaons, are here not
considered as projectiles. The reasons for this are
addressed in Discussion.

\section{Mean free path}

Imagine first a projectile is an electrically neutral rigid
ball ejected from an origin in a random direction, traveling
in a straight line. The volume of the cylinder it sweeps
after passing length $x$ equals $\uppi\rp^2 x$, where $\rp$
is the projectile's radius. Assuming a target particle is a
rigid ball of radius $\rt$ placed randomly in the sphere of
radius $x$, the probability that the projectile collides
with the target is ratio of volumes $\Vp/\Vtot$, where
$\Vp=\uppi(\rp+\rt)^2\,x$. The probability that the process
passes without collision is therefore 
\begin{equation}
q=1-\Vp/\Vtot
\label{eq:q}
\end{equation}
If the sphere contains two targets the probability equals
$(1-\Vp/\Vtot)^2$, and in case of $N$ targets the
probability is $(1-\Vp/\Vtot)^N$. As $N$ becomes large the
probability for survival equals
$$ 
p(x)=\lim_{N\to\infty}
\bigg(1-\frac{\uppi(\rp+\rt)^2\,x}{N/n}\bigg)^N
=\exp(-x/\lambda)
$$ 
where $1/\lambda=\uppi(\rp+\rt)^2\,n$ and volume density of
the target particles is $n=N/\Vtot$. For targets of different
types with radii $\rti$ and volume densities $n_i$, we find
\begin{eqnarray}
	\nonumber
	p(x)&=&\prod_{i}\lim_{N_i\to\infty}
\bigg(1-\frac{\uppi(\rp+\rti)^2\,x}{N_i/n_i}\bigg)^{N_i}=\\
&=&\prod_{i}\exp(-x/\lambda_i)=\exp(-x/\lambda) 
\label{eq:pxa}
\end{eqnarray}
where $1/\lambda_i=\uppi(\rp+\rti)^2\,n_i$ and
$1/\lambda=\sum_i(1/\lambda_i)$.

The mean free path of a proton in lower Earth atmosphere is
thus \req{pxa} 
\begin{equation}
\lambda_\Xp=\frac{1}{\sum_i\frac{1}{\lambda_i}}
	=\frac{1}{\uppi\,n\sum_i\eta_i(r_0+r_i)^2}
\doteq660~\mathrm{m}
\label{eq:lavg} 
\end{equation}
Classical values for proton radius $r_0=0.875$~fm and radii
of the target nuclei $r_i=r_0\,A_i^{1/3}$ with
$A_i=[14,16,40]$ were used in calculation of $\lambda_\Xp$.
For the volume density of the nuclei we assumed
$n=5.33\cdot10^{25}/\mathrm{m}^3$, and for the rates
$\mrm{\eta}{N}=78.39\%$, $\mrm{\eta}{O}=21.11\%$, and
$\mrm{\eta}{Ar}=0.502\%$.

\ignore{Distribution \req{pxa} is valid for electrically
neutral particles only. In such cases the values of
$\lambda$ and $\average{l}$ are equal.}

\section{Coulomb interaction}
\newcommand{\dd}[1]{\ensuremath{\mathrm{d}#1}}
\newcommand{\bpar}{\ensuremath{\kappa}}
\newcommand{\bpaq}{\ensuremath{\alpha}}


Coulomb interaction is responsible for the loss of
projectile's kinetic energy and ionization of its trajectory.
The rate of change in kinetic energy is expressed by Bethe
equation \cite{Bethe1934}. The range dependence for a proton
in lower Earth atmosphere is derived in \cite{Berkopec2016}
and shown on Fig.~\rep{range} for
$\lambda_{\Xp}\doteq660$~m. Since the height of a
thunder-cell is around 1~km or higher, graph on
Fig.~\rep{range} suggests the minimum initial energy for a
proton that reaches the ground is about 1~GeV.

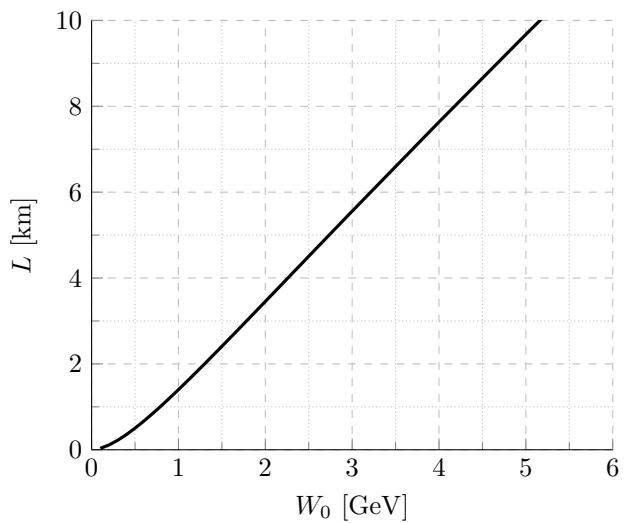
\begin{figure}[htb]
\centerline{
\begin{tikzpicture}[scale=1.0]
\pgfplotsset{major grid style={dashed}} 
\pgfplotsset{minor grid style={densely dotted}}
\begin{axis}[
	legend style={at={(0.68,0.36)},
		anchor=north west,draw=black,fill=white,align=left},
	legend cell align=left,
	grid=both,
	axis x line=bottom,
	axis y line=left,
	x axis line style={-},
	y axis line style={-},
	minor x tick num=1,
	minor y tick num=1,
	xticklabel pos=left,
	xtick align=inside,
	yticklabel pos=right,
	ytick align=inside,
	x tick label style={},
	ylabel={$L~[\mathrm{km}]$},
	xlabel={$W_0~[\mathrm{GeV}]$},
	domain=0.0:10.0,
	xmin=0.0,
	xmax=6.0,
	ymin=0,
	ymax=10,
	]
	\addplot[black,solid,very thick] table[x index=1,y expr=\thisrowno{9}*1e-3] {xavg_proton+1.dat};
\end{axis}
\end{tikzpicture}
}
\caption{Range $L$ for protons in lower Earth atmosphere 
as a function of their initial energy $W_0$.} 
\label{fig:range}
\end{figure}

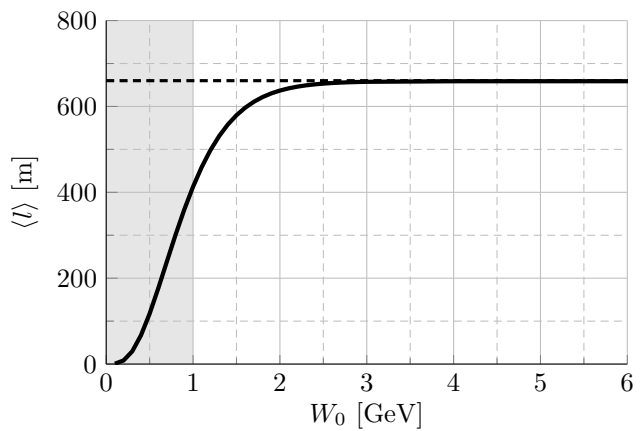
\begin{figure}[hbt]
\centerline{
\begin{tikzpicture}[scale=1.0]
\pgfdeclarelayer{background}
\pgfsetlayers{background,main}
\pgfplotsset{major grid style={solid}} 
\pgfplotsset{minor grid style={densely dashed}}
\begin{axis}[
	legend style={at={(0.60,0.52)},
		anchor=north west,draw=black,fill=white,align=left},
	legend cell align=left,
	grid=both,
	yscale=0.8,
	axis x line=bottom,
	axis y line=left,
	x axis line style={-},
	y axis line style={-},
	minor x tick num=1,
	minor y tick num=1,
	xticklabel pos=left,
	xtick align=inside,
	yticklabel pos=right,
	ytick align=inside,
	x tick label style={},
	xlabel={$W_0~[\mathrm{GeV}]$},
	ylabel={$\average{l}~[\mathrm{m}]$},
	domain=0.0:6.0,
	ytick={0,200,400,600,800},
	yticklabels={0,200,400,600,800},
	xmin=0.0,
	xmax=6.0,
	ymin=0,
	ymax=800,
	]
\begin{pgfonlayer}{background}
\fill[gray!20] (0,0) rectangle (1,800);
\end{pgfonlayer}
	\addplot[black!100,densely dashed,very thick] (0,660) -- (8,660);
	\addplot[black,solid,ultra thick] table[x index=1,y index=6] {xavg_proton+1.dat};
\end{axis}
\end{tikzpicture}
}
\caption{Average length $\average{l}$ between successive
	nodes (branching points) as a function of proton's
	kinetic energy $W_0$ at the first node. The dashed
	line represents estimation $\lambda_\Xp\doteq660$~m
	from \req{lavg}.} 
\label{fig:avgl}
\end{figure}
Estimation of the average length between two successive
nodes as a function of kinetic energy at the first node --
taking into account stopping power due to Coulomb
interaction -- was derived in \cite{Berkopec2016} and is
shown on Fig.~\rep{avgl}. Its impact on the mean free path
diminishes with increasing initial kinetic energy $W_0$.
Since protons with $W_0<1~\mr{GeV}$ do not reach the ground,
unless significant part of the projectiles has initial
energies $W_0$ in $[1~\mr{GeV}..2~\mr{GeV}]$ range, we may
assume Coulomb interaction can be neglected in the first
approximation.

\section{Proton-nucleus reactions}

Collision of \Xp~projectile with a nucleus $\X$ discussed so
far was assumed to be of the pick-up type
$\Xp+\X\to\X^{-}+\Xp+\Xp$, or $(\Xp,2\Xp)$, plus arbitrary
number of neutral particles. According to the hypothesis,
such reactions correspond to the observed binary-tree
geometry of lightning channels. Two types of reactions, swap
and capture, produce the results of collisions not accounted
for: swap can not be distinguished from a part of a branch
that has no split, and capture of a proton by a nucleus
looks like the end of a branch.

Now we extend the survival probability for a projectile and
one target \req{q} to reactions that are not necessarily of
pick-up type. Let us presume the pick-up reaction occurs
with probability $\ptp$. Then the channel does not split
with probability $q$ in case the target is not hit, or with
probability $(1-q)\cdot(1-\ptp)$ it the target is hit but
the reaction is not of pick-up type, or:
$$
q+(1-q)\cdot(1-\ptp)=
1-\ptp\frac{\Vp}{\Vtot}
$$
Assuming probability $\ptp$ is equal for all targets we find
for $N$ targets $(1-\ptp\Vp/\Vtot)^N$, and \req{pxa} changes
into $p(x)=\exp(-\ptp x/\lambda)$. From comparison with
\req{lavg} we see that for $\ptp<1.0$ the previous results
are valid after transformation
\begin{equation}
	\lambda\to\frac{\lambda}{\ptp}
	\label{eq:lpu}
\end{equation}

\section{Prediction of CG/CC ratio}

It is well documented that the ratio between the number of
strikes to the ground (CG) and the number of strikes that do
not reach the ground (CC) for a given storm depends on its
latitude. 

This dependence can be explained in view of our
hypothesis. We first presume that the average distance $R$
between the channel's origin in a thunder-cell and its most
distant point is independent of latitude. Second, we take
that the projectiles that reach the ground induce CG
lightnings while those that do not, induce CC lightnings. 

\begin{figure}[hbt]
\centerline{
\begin{tikzpicture}[scale=1.0]
\fill[black!10] (0,1) -- (1.73,0) -- (-1.73,0) -- cycle;
\fill[black!100] (-3,0) -- (3,0) -- (3,-0.06) -- (-3,-0.06) -- cycle;
\fill[black!40] (0,1) circle (0.20);
\node[anchor=west] at (2.5,0.5) {\small $h$};
\draw[stealth-stealth,ultra thin] (2.5,0) -- (2.5,1);
\draw[black!80,densely dashed] (0,1) circle (2.0);
\draw[black!40,dashed] (-0.5,1) -- (3,1);
\draw[stealth-stealth,ultra thin] (0,1) -- (-1.6,2.2);
\node[anchor=north east] at (-0.7,1.6) {\small $R$};
	\node[anchor=north] at (0,2.8) {\small CC};
	\node[anchor=south] at (0,0.2) {\small CG};
	\node[anchor=north] at (-2.4,0) {\small ground};
	\node[anchor=center] at (0.5,2.0) {\small thunder-cell};
\draw[-stealth,ultra thin] (0.4,1.8) -- (0.1,1.28);
\end{tikzpicture}
}
\caption{Spherical angle below thunder-cell (at the top of
	the shaded region) is proportional to the
	probability that a lightning strikes the ground.} 
\label{fig:cgcc}
\end{figure}
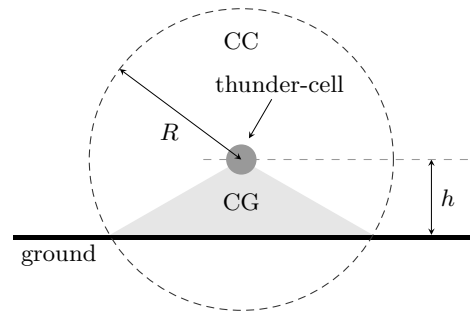

For projectiles ejected in a random direction the CG/CC
ratio can be estimated from the ratio of the areas defined
by the intersection of the sphere, whose center is a
thunder-cell at height $h$, and the plane, representing the
ground (see Fig.~\rep{cgcc}). The area of the sphere below
the plane correlates with incidence of CG lightnings
(corresponding spherical angle is shaded), the area above
the plane to CC lightnings. The ratio of the areas and
corresponding spherical angles equals
$$
\eta=\frac{R-h}{R+h}
$$
Taking the ratios $\eta_i$ and $\eta_j$ at two different
latitudes, where the heights of the thunder-cells are $h_i$
and $h_j$, respectively, one finds the estimation for
the ratio of thunder-cell heights reads
\begin{eqnarray}
\frac{h_i}{h_j}=\frac{1-\eta_i}{1+\eta_i}\cdot\frac{1+\eta_j}{1-\eta_j}
\label{eq:hihj}
\end{eqnarray}
Since a thunder-cell is always located close to the base of
its thundercloud we take that in the first approximation the
ratios of the heights $h_i/h_j$ and $H_i/H_j$ are close
enough $H_i/H_j\approx h_i/h_j$. At Equator at
$\latitude=\dg{0}$ experimental data gives $\eta_0=0.1$ and
$H_0\approx1100$~m. Together with \req{hihj} this leads to
prediction for CG/CC ratio from height of cloud-base:
\begin{equation}
\eta(H)=\frac{a-H/H_0}{a+H/H_0}
\label{eq:eta}
\end{equation}
where $a=(1+\eta_0)/(1-\eta_0)\approx11/9$.

\newcommand{\etaPM}{\ensuremath{\eta^{(\mathrm{a})}}}
\newcommand{\etaP}{\ensuremath{\eta^{(\mathrm{b})}}}

Based on empirical data Prentice and Mackerras
\cite{Prentice1977} for an estimate for CG/CC ratio suggest
\begin{equation}
\etaPM(\latitude)=\frac{1}{4.16+2.16\cos(3\latitude)}
\label{eq:etaPM}
\end{equation}
while Pierce \cite{Pierce1970} proposes 
\begin{equation}
\etaP(\latitude)=\bigg[\frac{1}{0.1+0.25\sin\latitude}-1\bigg]^{-1}
\label{eq:etaP}
\end{equation}

Values of CG/CC ratios for three latitudes from \req{etaPM}
and \req{etaP} along with our prediction \req{eta} are given
in Table \ret{lhe}.

\begin{table}[ht]
\centerline{
\begin{tabular}{|c|r||r|r|r|}
\hline
$\latitude$ & $H$~[m] & $\etaPM(\latitude)$ &
$\etaP(\latitude)$ & $\eta(H)$ \\
\hline
\hline
 $0^\circ$ &$1100$ & $0.16$& $0.11$ & $0.10$\\
\hline
$45^\circ$ & $700$ & $0.38$& $0.38$ & $0.32$\\
\hline
$60^\circ$ & $500$ & $0.50$& $0.46$ & $0.46$\\
\hline
\end{tabular}}
\caption{Height $H$ of a thunder-cloud base and CG/CC ratios
for three latitudes $\latitude$ (see \cite{Rakov2003}).
Values of $\etaPM$ and $\etaP$ follow from \req{etaPM} and
\req{etaP}. Our prediction \req{eta} is in column
$\eta(H)$.}
\label{tab:lhe} 
\end{table}

\section{Expected value of $\average{l}$}

Let us make a rough Fermi-type estimate about average length
$\average{l}$ from lightning photos and experience as
observers. We aim at higher confidence level and are
less concerned with error margin.

It is fairly safe to assume that less than 20\% of channels
have no nodes, and that no channel has more than 20 nodes.
Consequently, the remaining channels with number of nodes
between 1 and 19 occur with probability between 80\% and
100\%. If $M$ is the number of the nodes, the minimum and
the maximum expected values for $M$ are 
\begin{eqnarray*}
E(M)_\mr{min}&=&0\cdot0.2+1\cdot0.8+0\cdot20=0.8\\
E(M)_\mr{max}&=&0\cdot0.0+19\cdot1.0+0\cdot20=19
\end{eqnarray*}

Cloud-base heights range from $H_\mathrm{min}=500$~m to
$H_\mathrm{max}=1200$~m~\cite{Rakov2003}. It is impossible
to find the height of the cloud for a given lightning from
its photograph, so we take $L_\mathrm{min}\approx H$ because
the length of the main channel can not be shorter than the
minimum distance between the cloud base and the ground,
while the maximum length is taken to be five times that, or
$L_\mathrm{max}\approx5\cdot H$ (in such case the average
direction of the channel is \dg{80} from vertical, and the
lightning strikes the ground around 4.8~km from the cloud). 

The lower and upper expected values for the average length
between successive nodes are then
\begin{eqnarray*}
E(\average{l})_\mathrm{min}&=
\frac{L_\mathrm{min}}{E(M)_\mathrm{max}+1}
=\frac{500}{19+1}&=25~\mathrm{m}\\
E(\average{l})_\mathrm{max}&=
\frac{L_\mathrm{max}}{E(M)_\mathrm{min}+1}
=\frac{5\cdot1200}{0.8+1}&\approx 3.3~\mathrm{km}
\end{eqnarray*} 
\ignore{The expected value can not exceed 1311.2~m, as
follows from \req{pxa} and \req{lavg} for a projectile of
radius \rp=0~m.}
The range spans over three orders of magnitude and includes
our prediction of 660~m for protons and $(\Xp,2\Xp)$
reactions. For longest expected value
$\average{l}\approx3333~\mr{m}$ the probability $\ptp$
according to \req{lpu} equals
$\ptp\approx660/3333\approx20\%$. 

\section{Discussion}

Estimation of $\average{l}$ above has a wide error margin
but the particles that may be involved in the process do not
come in arbitrary sizes. Classical prediction from
\req{lavg} for projectile of zero size gives
$\lambda\approx1311$~m, and charged particles larger than
proton are either composite or short-lived. 

Secondary projectiles from proton-nucleus reactions often
include pions and kaons. Their classical radii give
$\lambda_{\Xpi}\doteq700$~m for pions and
$\lambda_{\XK}\doteq810$~m for kaons. Short-lived pions
decay in muons, and muons decay in positrons or electrons,
so these can only contribute a non-branched part to a
channel. One specific decay of kaons $\XKpm\to3\Xpipm$ may
lead to channel branching but it is less likely to occur as
kaons themselves are the rarest secondary products among the
$\Xp$, $\Xpipm$, and $\XKpm$, and since this type of decay
for kaons has only 6\% rate.~\cite{pdg2012booklet}

Experimental verification of the hypothesis should involve
measurements of the average length between successive nodes
by reconstruction of 3D channel geometry. Correlation
between intensity of freezing/precipitation and frequency of
lightning is expected, as well as the ejection of elementary
particles from super-cooled water during freezing. 

\ignore{could be
determined by measurements of lightning geometry and
detection of elementary particles ejected from a
super-cooled water during freezing.}


\begin{thebibliography}{11}

\bibitem{Gunn1948}
R.~Gunn,
\newblock {Electric field intensity inside of natural
clouds}. \newblock {\em Journal of Applied Physics},
19(5):481--484, 1948.

\bibitem{Marshall1991}
T.C.~Marshall and W.D.~Rust,
\newblock {Electric field soundings through thunderstorms}.
\newblock {\em Journal of Geophysical Research: Atmospheres},
{96}(D12):{22297--22306}, {1991}.

\bibitem{Winn1974}
{W.P.~Winn et al.},
\newblock {Measurements of electric fields in thunderclouds},
\newblock {\em Journal of Geophysical Research},
{79}({12}):{1761--1767}, {1974}.

\bibitem{Berkopec2012}
	{A.~Berkopec},
	\newblock{Fast particles as initiators of stepped
	leaders in CG and IC lightnings},
\newblock {\em Journal of Electrostatics},
70(5):462--467, 2012.

\bibitem{Scott2014}
{C.J.~Scott et al.},
\newblock{Evidence for solar wind modulation of lightning},
\newblock{\em Environmental Research Letters}, 9(5), {2014}.

\bibitem{Berkopec2016}
	{A.~Berkopec},
	\newblock{About Geometry and Initial Phase of 
	Cloud-to-Ground Lightning},
	\newblock{arXiv:1602.02496},
	{physics.ao-ph}, {2016}.

\bibitem{Bethe1934}
H.~Bethe and W.~Heitler,
\newblock On the stopping of fast particles and on the
creation of positive electrons.
\newblock {\em Proceedings of the Royal Society of London. Series A},
146(856):83--112, 1934.

\bibitem{Rakov2003}
{V.A.~Rakov and M.A.~Uman},
\newblock{Lightning: physics and effects}, {2003},
{Cambridge University Press}.

\bibitem{Prentice1977}
	{S.A.~Prentice, D.~Mackerras},
	\newblock{The ratio of cloud to cloud-ground
	lightning flashes in thunderstorms},
	\newblock{\em J.~Appl.~Meteor.},
	545--550, 1977.

\bibitem{Pierce1970}
	{E.T.~Pierce},
	\newblock{Latitudinal variation of lightning
	parameters},
	\newblock{\em J.~Appl.~Meteor.},
	194--195, 1970.

\bibitem{pdg2012booklet}
J.~Beringer et al. (Particle Data Group),
\newblock {\em Phys. Rev. D86}, 010001, 2012.

\end{thebibliography}
\end{document}